# 2D semiconductors as integrated light sources for plasmonic waveguides

Christian Frydendahl*[a], Torgom Yezekyan[a], Vladimir Zenin[a], Sergey I. Bozhevolnyi[a]
[a]University of Southern Denmark, Centre for Nanooptics, 55 Campusvej, Odense, DK 5230

## ABSTRACT

2D materials are compatible with many material platforms as they adhere to other materials strictly by van der Waals interactions. This, together with the variety of band gaps found among transition metal dichalcogenides (TMDCs), makes them uniquely interesting to study for on-chip light sources, as the photonic integrated circuit (PIC) industry is developing into several material platforms depending on the application (gold, silicon, silicon nitride, indium phosphide, etc). 2D materials could thus potentially provide a universal on-chip light source (and detector) scheme for all PIC platforms in the future. Here we show recent results on how exfoliated monolayer TMDC flakes can be integrated with plasmonic slot-waveguides and plasmonic dipole antenna couplers to inject their photoluminescence into plasmonic waveguides as a first step towards future on-chip light sources for such systems.

**Keywords:** Plasmonic waveguides, 2D materials, nanophotonics, photoluminescence, optical antennas

## 1. INTRODUCTION

Integrated photonic systems are promising to enable the miniaturization of optical systems and measurements to chip-scale devices[1-2], with applications in new quantum devices[2], ultra high-speed telecommunications platforms[3], diagnostics[4], new computing architectures[5], and more. Integrated systems based on plasmonic waveguiding is particularly interesting for sensing and high-speed telecommunications applications, where the ability of plasmonic confinement of light below the diffraction limit enables single molecule sensing and major reductions in device capacitance for optoelectronic modulators due to greatly reduced device footprint[3].

However, establishing the connection between integrated optics and the outside world remains the biggest challenge for widespread commercialization and adoption of photonic integrated circuits (PICs). To inject light into integrated systems, it must be coupled from the outside with a lensed fiber or microscope objective through an on-chip coupling structure (e.g., a grating), limiting the scalability and ultimate promise of independent optical chips. Making direct, on-chip, light sources that directly inject light into the chip is thus a crucial step for delivering on the many promising applications of PICs. However, the material of choice for the waveguide component of the chip is specifically picked due to its waveguiding capabilities, i.e., lack of absorption/emission. Therefore, another material must generally be introduced onto the same photonic chip to play the part of the light source or detector.

2D materials are compatible with many material platforms as they adhere to other materials strictly by van der Waals interactions. This, together with the variety of band gaps found in TMDCs (and their potential tunability), makes them uniquely interesting to study for on-chip light sources, as the PIC industry is developing into several material platforms depending on the application (gold, silicon, silicon nitride, indium phosphide, etc). 2D materials could thus provide a universal on-chip light source (and detector) scheme for all PIC platforms in the future.

Here we demonstrate how the emitted photoluminescence of the TMDC MoSe$_2$ can be coupled directly into a plasmonic slot-waveguide mode by placing the material flake at the input facet of the waveguide, such that plasmonic antennas on the same chip stimulate emission into the waveguide upon external illumination.

*cfry@mci.sdu.dk

## 2. RESULTS

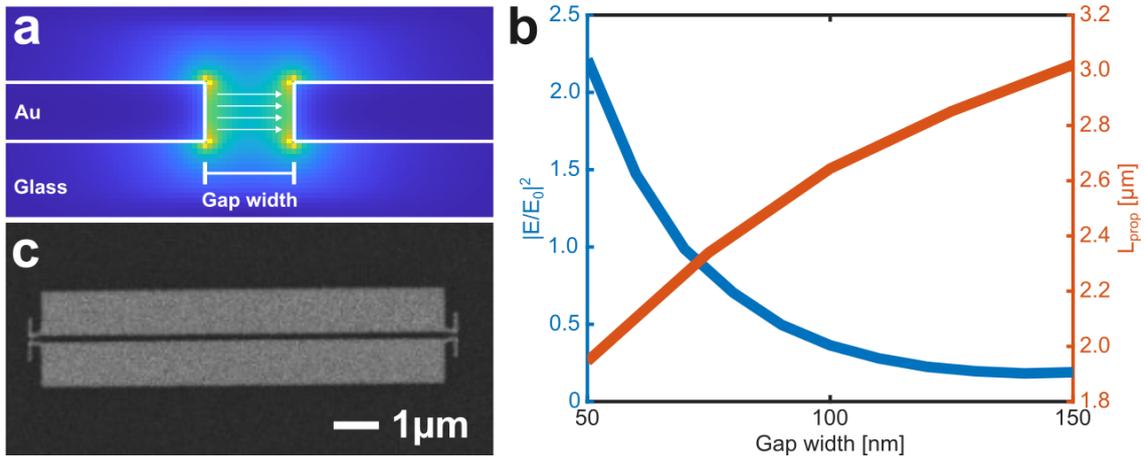

Figure 1. **a)** Cross-section of the field associated with the plasmonic slot mode. **b)** Results from Optical simulations showing the field enhancement attained in a nanoantenna gap vs. the antenna gap width, and the propagation length for the fundamental mode of a plasmonic slot-waveguide vs. the gap width. **c)** Electron micrograph of a fabricated waveguide.

For our waveguide design we choose plasmonic slot-waveguides as these have most of the electric field of their fundamental mode aligned parallel to the substrate surface (TE mode). This is exactly the field orientation required to couple to the radiative excitons in $MoSe_2$ monolayers, and as such we expect to see efficient coupling between excited photoluminescence from the $MoSe_2$ and the propagating mode of the waveguides. To excite the mode in the structure we have opted to use plasmonic gap antennas[1]. These antennas allow for a an extremely small coupling footprint, while providing a very localized plasmonic hotspot where we expect to see enhanced emission from the $MoSe_2$ when positioned inside the antenna gap.

First, we simulate the waveguide and antenna structures using Lumerical FDTD to find the best parameters for optical excitation, propagation, and which are also realistic for fabrication, Fig. 1.a. We find the best parameters to be 100 nm thick metal films with a gap/slot width that tapers from 50 to 150 nm over a 250 nm distance. In this way, we can excite the input antenna couplers with the minimum 50 nm gap, and then quickly expand the mode to the 150 nm gap width to increase the propagation length, Fig. 1.b. Next, we fabricate several waveguides with both input and output antenna couplers at the ends, Fig. 1.c, using standard electron-beam lithography and thermal gold evaporation.

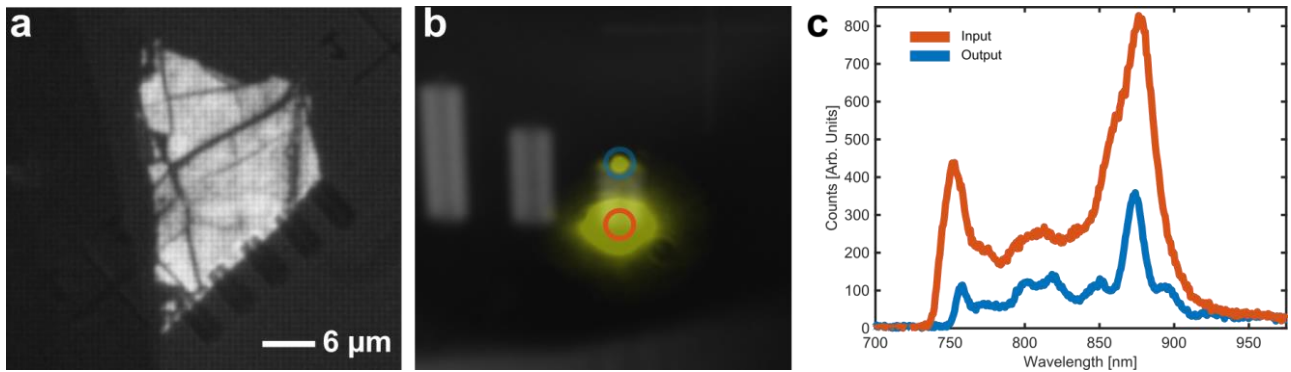

Figure 2. **a)** Photoluminescence image of $MoSe_2$ layer transferred to the input couplers of plasmonic waveguides (dark outlines). **b)** Optical image of excitation of a single waveguide with $MoSe_2$ transferred onto one set of antenna couplers. **c)** Optical spectra measured from the input and output regions of the device, showing the same optical signal is transmitted from the input the output. The peak at 750 is near the cut-off of the long-pass filter used to cut the excitation laser.

Next, we exfoliate few and monolayer MoSe$_2$ flakes onto PDMS from bulk crystals using Nitto dicing tape. We then identify mono- and bilayers by optical contrast microscopy, and confirm that the exhibit efficient photoluminescence in a custom-built photoluminescence imaging system, Fig. 2.a. Then, using a standard dry visco-elastic transfer process and a custom transfer system, we pick up the exfoliated flakes and transfer them to the antenna couplers of the plasmonic waveguides. We perform the drop-down of the flakes in the transfer step by melting off a sacrificial polycarbonate (PC) layer. In this way, the transferred flakes are automatically encapsulated in a layer of PC, protecting them from the environment to some degree, and also matching the refractive index above and below the gold layers of the waveguides. If there is not index matching layer, the plasmonic mode would not be confined to the narrow gap (instead simply leaking into the glass substrate).

After this transfer, we can now excite the photoluminescence of the MoSe$_2$ with a laser polarized to match the plasmonic antennas, thus exciting the photoluminescence directly into the waveguides. By filtering out the excitation laser with a long-pass filter, we can then see the photoluminescence couple out at via the output antenna couplers, Fig. 2.b. To confirm that the emission is the same at the input and the output, we perform spectroscopy of both areas independently, using an aperture to spatially filter only the signal from either area of the device, Fig. 2.c.

### 3. CONCLUSIONS

We have demonstrated how the light emitted from a TMDC can be coupled directedly into a guided mode in a plasmonic waveguide. Our result is a first step in the process of making on-chip light sources using 2D materials, with the next planned step to demonstrate coupling of electroluminescence from a 2D material heterostructure junction.

### ACKNOWLEDGEMENTS

C.F. is supported by the Carlsberg Foundation as a Carlsberg Reintegration Fellow (grant number: CF21-0216).